# Effects of thermal rippling on the frictional properties of free-standing graphene

A. Smolyanitsky

Applied Chemicals and Materials Division,
National Institute of Standards and Technology
Boulder, CO 80305
alex.smolyanitsky@nist.gov

**Abstract**

With use of simulated friction force microscopy, we present the first study of the effect of varying temperature on the frictional properties of suspended graphene. In contrast with the theory of thermally activated friction on the dry surfaces of three-dimensional materials, kinetic friction is demonstrated to both locally increase and decrease with increasing temperature, depending on sample size, scanning tip diameter, scanning rate, and the externally applied normal load. We attribute the observed effects to the thermally excited flexural ripples intrinsically present in graphene, demonstrating a unique case of temperature-dependent dynamic roughness in atomically thin layers. Consequently, our results suggest strain-induced control of friction in nanoelectromechanical systems involving free-standing regions of atomically thin membranes.

**Introduction**

The use of free-standing and substrate-bound atomically thin layers in electronics and nanomechanical devices requires understanding of their interfacial properties, which includes kinetic friction. Not surprisingly, friction at the nanoscale has recently joined the list of interesting phenomena observed for these materials. Among key results, friction force microscopy (FFM) measurements on epitaxial few-layer graphene revealed extremely low kinetic friction, which suggested use of graphene as a solid-state lubricant [1-5]. Experimental studies [6] and numerical simulations [7-9] demonstrated a distinct reduction of kinetic friction with increasing number of suspended stacked layers. Highly nonlinear friction-load dependencies were predicted for free-standing [10] and experimentally observed for substrate-bound and free-standing few-layer graphene [11]. In addition, increasing friction with decreased applied load was reported for the oxidized surfaces of lamellar materials [12-14]. Finally, the effects of surface roughness on the frictional properties of fluorinated [15] and hydrogenated[16] graphene were reported.

The majority of studies to date have focused on the structural properties, including the effects of impurities and graphene-substrate interactions. The dependence of friction on temperature in the case of atomically thing layers, of utmost importance for technological applications, is currently unknown. Here, we present a carefully designed molecular dynamics (MD) simulation study of the effects of varying temperature on the kinetic friction observed on locally suspended graphene, as described in Fig. 1. In particular, we demonstrate the effect of thermally excited dynamic ripples [17, 18] on the frictional properties of free-standing graphene laterally scanned by a simulated FFM tip, presenting a case of dynamic roughness in atomically thin membranes.

**System description**

The systems simulated here employed the geometry presented earlier [8, 10], shown in Fig. 1. The main simulated graphene sample consisted of 8192 atoms (13.6 nm × 15.8 nm);

additional samples consisted of 5408 atoms (11.1 nm × 12.8 nm) and 1352 atoms (5.5 nm × 6.4 nm). The tips were modeled by (5,5) and (10,10) capped single-wall carbon nanotubes (SWCNTs) with effective tip diameters $d = 1.2\ nm$ and $2.2\ nm$, respectively. All graphene samples were statically pre-relaxed with use of the optimized second-generation bond-order potential [19]. The scans were initiated by a rigidly translated virtual cantilever connected to the upper part of the scanning tip via lateral springs. The effective lateral stiffness was 10 N/m, in accord with previous simulations and consistent with the typical lateral compliance of FFM cantilevers[9, 16, 20]. The cantilever's constant scan velocity was $v = 1\ m/s$, unless stated otherwise. Weak integral feedback control was imposed on the tip's vertical position to maintain average prescribed normal loads, mimicking a constant-load FFM scan. Previously parameterized [8] thermal control via Langevin thermostat [20] was applied to the boundary regions, as shown in Fig. 1. All tip-sample interactions were simulated via Lennard-Jones potential [21] with $\varepsilon = 7.5$ meV and $\sigma = 0.31$ nm, resulting in a flat-on-flat graphene interlayer adhesion strength of $42.8\ \frac{meV}{atom}$, in agreement with *ab initio* and experimental data for graphite [22-24]. The intramolecular interactions within the tip and the graphene layer were maintained with use of the computationally efficient bond-order [19] informed harmonic model [25], shown to reproduce the structural properties of graphene and carbon nanotubes [26] and demonstrate reasonable accuracy in the qualitative description of thermal fluctuations in graphene, as compared with the bond-order potential [27]. See Supplemental information for further detail on the computational methods and calculations of friction forces (time-averaged lateral forces experienced by the tip in the direction of the scan). Unless stated otherwise, the simulated time was 10 ns.

**Results and discussion**

Shown in Fig. 2 (a) is a set of friction forces as functions of increasing temperature normalized with respect to the friction force at the lowest temperature of 2 K at various normal loads for a 8192-atom graphene sample. For the lowest normal load of 0 nN, the average friction force increases considerably throughout the entire simulated temperature range. The relative effect decreases with increasing normal loads with friction force very slightly decreasing at T > 250 K in the 6 nN load curve, qualitatively starting to approach the experimentally supported behavior for graphite [28]. From the classical standpoint of the Tomlinson model [29], as well as from the modern view of thermally-activated friction [30], the amount of sliding friction *decreases* with increasing temperature as a direct result of thermal "smoothing" of the periodic tip-sample energy profile (see Fig. 3 (b)). Consequently, the effective lateral force peak value *decreases*, along with the average friction force offset. Therefore, the $\left(\frac{dF}{dT}\right) > 0$ (where F is the average friction force) trends are generally not expected for the dry surfaces of three-dimensional solids, as supported by experiment,[28, 31] although possible for a wet contact [32] and in humid environments [33]. Friction increasing within a short temperature range around 100 K was previously reported for the dry silicon-silicon contact [31]. Here, however, the amount of friction is observed to increase *in the entire simulated temperature range*. A possible contribution to the observed effect is from the temperature-varying tip indentation depth, which effectively controls the viscoelastic contribution to friction [8, 10]. However, given the data in Fig. 2 (b), where we show the out-of-plane deformation profiles of the 8192-atom graphene sample presented in Fig. 2 (a) (obtained from mapping the atomic position snapshots on a grid along the scan vector), such a contribution is unlikely. For the temperatures considered, the largest indentation depth in Fig. 2 (b) is in fact at the *lowest* temperature of 2 K, as can be expected from membrane's self-stiffening as the temperature increases. This effect is due to nonlinear coupling between the in- and the out-of-plane vibrational modes, resulting in an increase of the bending rigidity of the membrane at finite temperatures [34]. At the normal load of 0 nN in Fig. 2 (c), where the effect of friction increasing with temperature is most greatly pronounced, the differences are only in the flexural corrugation of the sample at higher temperatures. The high crystalline order of the small scanning tips at all simulated temperatures also suggests that the idea of temperature-dependent individual contacts formed between an amorphous tip and the sample surfaces [35] proposed to explain the previous experimental observation of local $\left(\frac{dF}{dT}\right) > 0$ trends [31] is not applicable in our case.

In addition to free-standing samples, we studied the dependence of friction on temperature in the case of a 1352-atom sample, in which all atoms were harmonically restrained to their initial positions in the flat phase (left inset of Fig. 1), effectively representing a sample strongly supported by substrate. The results are shown in Fig. 3 (a), where we observe $\left(\frac{dF}{dT}\right) < 0$ trends similar to the experimental data on graphite [28] and in agreement with existing theory [29, 30]. Simulations of the same sample performed at a lower scan rate of 0.1 m/s revealed identical decreasing trends.

It is noteworthy that high simulated scan rates can have a profound effect on the friction's dependence on temperature, such as that in Figs. 2 (a) and 3 (a). According to the existing theory, the dependence of friction on temperature is ultimately affected by the so-called critical sliding velocity [30], applicable in our case, because the lateral force profile remains periodic (see Section S1 of the Supplement). This threshold velocity separates the regimes of stick-slip and continuous sliding along a periodic energy profile, and is equal to $v_0 \cong \frac{\pi f_0 k_B T}{\sqrt{2} k a}$, where $f_0$ is the maximum lateral attempt frequency (a fitting parameter in this case), $k_B$ is the Boltzmann constant, $a = 0.246\ nm$ is the lattice constant of graphene, and $k$ is the tip-sample lateral stiffness [30]. With $k \cong 1.5\ N/m$ (from the typical lateral force as a function of the lateral sliding distance) and $f_0 \approx 100$ GHz, the critical velocity limits for the suspended and supported samples at T = 2 K and 500 K are 2 cm/s and 4.2 m/s, respectively. The value of $v_0$ reaches 1 m/s at T = 120 K, which suggests that the simulated scans at 1 m/s were performed in the continuous sliding regime at T < 120 K (see Supplemental Information). Despite the $v_0 \sim T$ dependence, theory predicts $\left(\frac{dF}{dT}\right) < 0$ for *the suspended and the supported samples* in the entire simulated temperature range (see Supplemental Information). We therefore have an outstanding qualitative difference observed between the free-standing and the supported samples at low normal loads, which cannot be explained within the current theoretical model. The observed difference therefore suggests an applicability limit of the existing theoretical model for the free-standing atomically thin membranes. Importantly, as we show further, lateral scans performed at 0.1 m/s yield a friction *vs.* temperature dependence similar to that shown in Fig. 2 (a). We therefore believe that a different, more general mechanism is responsible for the differences observed between the free-standing and supported samples.

We propose that the behavior observed in Fig. 2 (a) is a direct effect of the tip interacting with the thermally excited dynamic ripples intrinsic for free-standing atomically thin membranes [17, 18]. Atomically thin crystals are fundamentally barred from strict two-dimensionality at finite temperatures [17, 18, 34, 36, 37], and thermally excited flexural ripples were recently measured with use of scanning tunneling microscopy (STM) [38]. Such ripples effectively present dynamic random asperities in front of the moving tip. The effect is not present in supported samples, because the ripples are effectively suppressed, allowing only for the high-frequency atomic vibration around the equilibrium sites. As a result of rippling in free-standing samples (as clearly observed in Fig. 2 (c)), during lateral scanning, the tip is probing a *dynamically corrugated surface,* as shown schematically in Fig. 3 (b) and also directly visualized from our simulations in Fig. 3 (c). It is important to realize that the dynamic rippling effects considered here are separate from (and additional to) the *quasistatic* wrinkles induced in atomically thin layers by strain, or boundaries, [39-41] also structurally expected from considerably thicker polymer layers [42].

According to the theory of thermally fluctuating membranes, the ripples' Fourier amplitudes $h(q)$ scale with the membrane temperature $T$ [17, 34] as

$$\langle h^2(q) \rangle = \frac{k_B T}{\kappa} \frac{1}{q^4}, \qquad (1)$$

where $\kappa$ is the membrane's bending rigidity. Eq. (1) implies $h(q) \propto 1/q^2$ and yields the following average [17, 34]:

$$\langle h^2 \rangle \cong \frac{k_B T}{8\pi^2 \kappa} L^2, \qquad (2)$$

where $L$ is the characteristic membrane size. For the membrane dimensions considered here, both theoretically calculated and simulated root-mean-square heights $\sqrt{\langle h^2 \rangle}$ are non-negligible compared to the 1.42 Å C-C bond length (see Supplemental Information), in accord with previous reports on graphene membranes of similar dimensions [17, 18, 37]. Further, the STM measurement yields $\sqrt{\langle h^2 \rangle} \cong 1.4\ nm$ for a considerably larger free-standing membrane [38], suggesting a possibly significant effect of ripples in experimental FFM scans. Such rippling magnitudes suggest that free-standing graphene cannot be assumed to be atomically flat during a FFM scan at normal loads sufficiently low not to suppress the rippling process via membrane stretching.

It is straightforward to demonstrate that for a static corrugated surface, according to the Tabor model [43], the friction force $F$ increases with the effective roughness as

$$F = F_0(T)(1 + \varepsilon\langle\varphi^2\rangle), \qquad (3)$$

where $F_0$ is the friction force on an atomically flat surface subject to thermal activation, $\varepsilon$ is a dimensionless coefficient corresponding to the degree of plasticity in the tip-sample interactions, and $\varphi$ is the tip-asperity effective incidence angle assumed to be small. Regardless of the dynamic nature of the ripples considered here, a similar discussion is possible at least in the qualitative sense due to the overall statistical effect the of ripples' presence, which strongly contributes to the tip-sample energy dissipation during a FFM scan. Note that Eq. (3) is qualitatively valid only at normal loads close to zero, as it generally neglects the viscoelastic frictional component from sample indentation, described earlier for higher loads in smaller samples [8], where the rippling process was suppressed and thus conventional frictional mechanisms dominated. Also, the additive form of Eq. (3) in this case is unlikely to capture the complete physical picture, because the presence of ripples affects *the thermal activation mechanism itself*, as discussed below. Nevertheless, a qualitative discussion is possible. Let us consider an additive correction to the conventional thermally activated van der Waals bonding-debonding process (corresponding to the results in Fig. 3 (a)) so that total friction takes the form of Eq. (3). In this case, $\varepsilon$ is a dimensionless parameter depending on the normal load, scan velocity, and possibly boundary conditions (including the boundary dissipative properties), and $F_0(T)$ is the thermally activated friction force, which, as we show further, should include the effect of rippling. For a point slider, $\varphi$ is the angle between the local normal and the *001* direction [43], while for a tip of finite diameter $d$, $\langle\varphi^2\rangle \propto 4\sqrt{\langle h^2\rangle}/d$. Substituting the latter expression with use of Eq. (2) into Eq. (3), we obtain

$$F(T) \cong F_0(T)\left[1 + \frac{\varepsilon L}{\pi d}\sqrt{\frac{2k_B T}{\kappa}}\right]. \qquad (4)$$

Approximating the thermally activated mechanism as $F_0(T) = F_{0K} - \Delta F(T)$ (*e.g.* Fig. 3 (a)), the normalized values in Fig. 2 are given by

$$\frac{F(T)}{F_{0K}} \cong (1 - b(T))\left[1 + \left(\frac{\varepsilon L}{\pi d}\right)\sqrt{\frac{2k_B T}{\kappa}}\right] \qquad (5)$$

with $b(T) = \frac{\Delta F(T)}{F_{0K}} \in [0|_{T=0K}, 1)$, describing an effective competition between the product terms. Equation (5) reduces to the standard thermally activated model for large $\kappa$ (corresponding effectively to the surfaces of three-dimensional solids). It also suggests a $\sqrt{T}$ dependence for $b(T) \ll 1$, similar to the results in Fig. 2 (a). In addition, within Eq. (5) the observation of increasing load-induced gradual return to the conventional behavior can be explained by ripple suppression due to membrane stretching due to external loads and any boundary-induced pre-strain, which affects both $\kappa$ and $b(T)$. A more subtle point should be made regarding the effect of finite tip size even within the simplified view of Eq. (5). A finitely-sized round slider of diameter $d$, as is often the case for FFM tips, is ripple-selective beyond the *1/d* dependence in Eq. (5). The ripples with wavelengths below $d$ are not expected to present considerable asperities *in front of the tip*. Instead, such short ripples will be located *under the tip* (see Fig. 3 (b)), contributing to the thermally activated component $F_0(T)$. Therefore, the shortest ripple wavelength capable of introducing a randomly present asperity in front of the tip is $d$, for $d \ll L$ resulting in a further tip size effect:

$$\frac{F(T)}{F_{0K}} \cong (1 - b(T)) \left[1 + \left(\frac{\varepsilon L}{\pi d}\right)\sqrt{\frac{2k_B T}{\kappa}}\left(1 - \frac{d^2}{2L^2}\right)\right]. \quad (6)$$

The effect of short ripples under the tip is expected to be significant for the thermally activated mechanism itself, even if we neglect the strong Bragg peaks in the short-wavelength part of the distribution (see Supplemental Information). The overall amount of rippling *under the tip* from Eq. (1) is $\langle h_{short}^2 \rangle \cong \frac{k_B T}{8\pi^2 \kappa}(d^2 - a^2) \cong 0.02 \text{ Å}^2$ for T = 300 K, $d = 1.2 \text{ nm}, \kappa = 2.2 \text{ eV}$ [27] – an order of magnitude larger than the Debye-Waller factor of graphene [44, 45], critical to the thermal activation mechanism. The overall effect of rippling can therefore be summarized as follows: increasing temperature causes the presence of long-wave asperities of increasing height in front of the tip (via the $\left[1 + \left(\frac{L}{d}\right)\sqrt{T}\right]$ term), while each asperity becomes more "slippery" via short ripple -enhanced thermal activation (via the $(1 - b(T))$ term).

The discussion above leads us to an important point: in deriving Eq. (6), we utilized Eq. (1) from the classical theory of thermally fluctuating membranes, resulting in the linear $\left(\frac{L}{d}\right)$ scaling of the $\sqrt{T}$ term. In reality (at least partially represented by an atomistic model), the height distribution of ripples can strongly depart from the $1/q^2$ scaling in Eq. (1) due to coupling between stretching and bending modes [17, 18] (also see Supplemental Information). Therefore, given the competitive nature of Eq. (6), one can expect a far more intricate dependence of the low-load friction on the sample size, and/or local boundary-induced lateral strains via modification of the ripple height distribution, resulting in a significant modification to the entire long-ripple contribution. Of particular importance, the linear scaling with the effective membrane size $L$ may have to be replaced with $L^{(1-\eta/2)}$ where η is an effective, possibly temperature (and size) dependent fitting parameter described elsewhere [18], especially relevant to membranes considerably larger than those discussed here. Therefore, as stated above, significant changes to both the $\sqrt{T}$ and *b(T)* terms in Eq. (6) are contributed to by the ripples in front of and under the tip, respectively (also see section 4 of Supplemental Information).

In addition to the discussion above, the effect of tip diameter in Eq. (6) for a given 8192-atom sample can be observed in Fig. 4 (a), where we compare friction vs. temperature, as obtained with tip diameters of 1.2 nm and 2.2 nm. For the *d* = 2.2 nm tip, the normalized friction values are *lower* than those obtained with a *d* = 1.2 nm tip, consistent with our earlier suggestion that the tip diameter scaled $\sqrt{T}$ term in Eq. (6) dominates the results in Fig. 2 (a). In general, we believe that the effects described here are observable for larger tips, *provided that FFM scans are performed on L ≫ d membrane samples* due to membrane size scaling. Interestingly, the sensitive dependence of friction on the rippling distribution and the tip size suggests a possibility of externally excited ripples [37, 38], as well as strain-induced modification of the rippling patterns for controlling friction in nanoelectromechanical applications, possibly including controlled superlubricity.

Experimental observability of the rippling effects at considerably lower lateral scan rates is important, provided the vast difference between experimental FFM scan rates and those used in MD simulations. We therefore performed additional simulations at considerably lower scan rates of 0.1 m/s (simulated for 40 ns) and 0.25 m/s (simulated for 20 ns). We selected the 5408-atom sample, which exhibits a $\left(\frac{dF}{dT}\right) < 0$ trend during a 1 m/s scan. As shown in Fig. 4 (b), local $\left(\frac{dF}{dT}\right) > 0$ regions are observed at lower scan rates. The changes in monotonicity of *F(T)* again suggest a strong velocity dependence in the competing terms in Eqs. (5,6). Our results therefore indicate a possibility of observing the effects of thermally induced ripples on friction at lower lateral scanning rates.

**Conclusions**

Our work demonstrates a unique case of temperature-dependent dynamic roughness observed in suspended graphene as a result of thermally excited flexural waves. In contrast with the existing theory and experimental results for the surfaces of ordinary solids, we demonstrate the possibility of kinetic friction to both increase *and* decrease with increasing temperature, depending on the tip size, scanning rate, and lateral strain imposed upon the membrane sample. The effects reported are likely be observable experimentally and can occur under low normal loads in other free-standing atomically thin membranes. The sensitive dependence of friction to the flexural ripple patterns suggests control of frictional properties of atomically thin membranes (including imposed superlubricity) via externally excited flexural vibrations, as well as via externally applied lateral strains.

**Acknowledgment**

The author is grateful to R.J. Cannara, J.P. Killgore, A.F. Kazakov, V.K. Tewary, and K. Kroenlein for useful discussions and comments. This work is a contribution of the National Institute of Standards and Technology, an agency of the US government. Not subject to copyright in the USA.

**Figures**

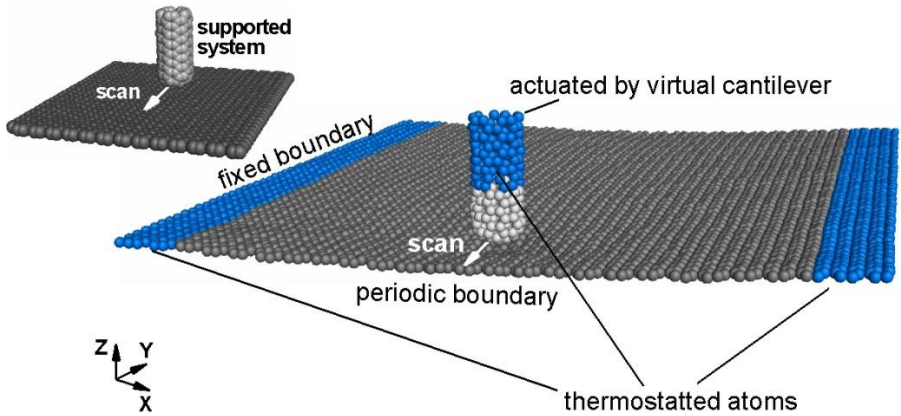

Figure 1. Simulation setup depicting the thermal control, boundary conditions, and the direction of the scan. In the supported graphene sample (inset on the left) the boundaries and the thermostatted regions are set up identically to the suspended samples, while all atoms in this case are harmonically restrained against displacement.

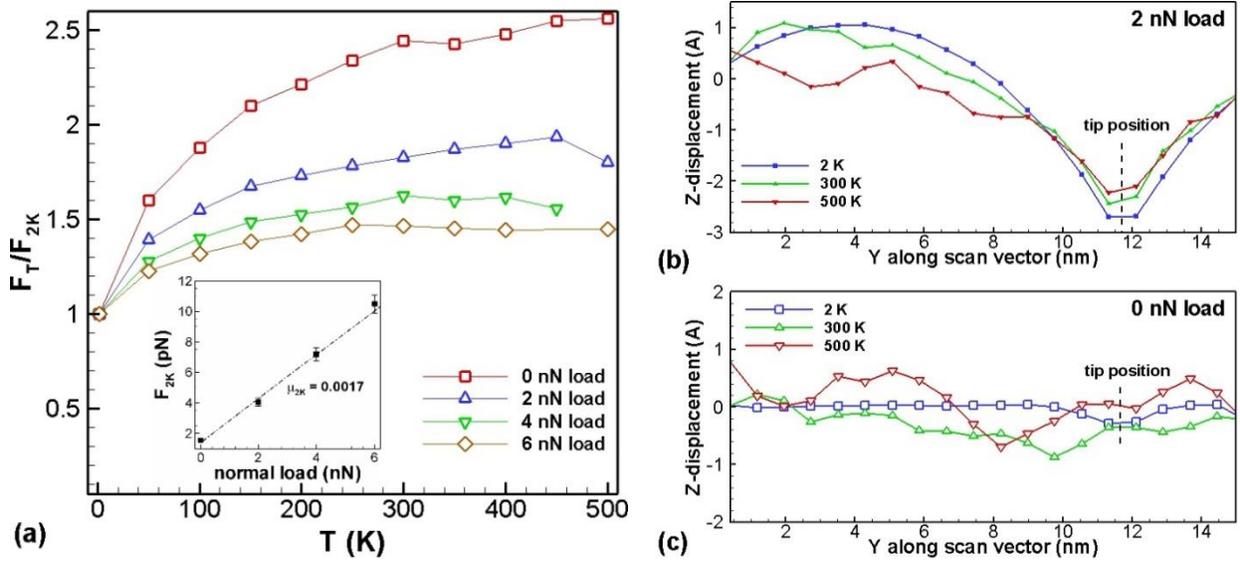

Figure 2. Normalized (with respect to the value at lowest T = 2 K) average force as a function of temperature at various normal loads for a suspended monolayer graphene sample consisting of 8192 atoms (a) and out-of-plane deformation profiles along the simulated lateral scan vector for various temperatures and normal loads of 2 nN (b) and 0 nN (c). The inset in (a) shows the friction force at 2 K as a function of the normal load.

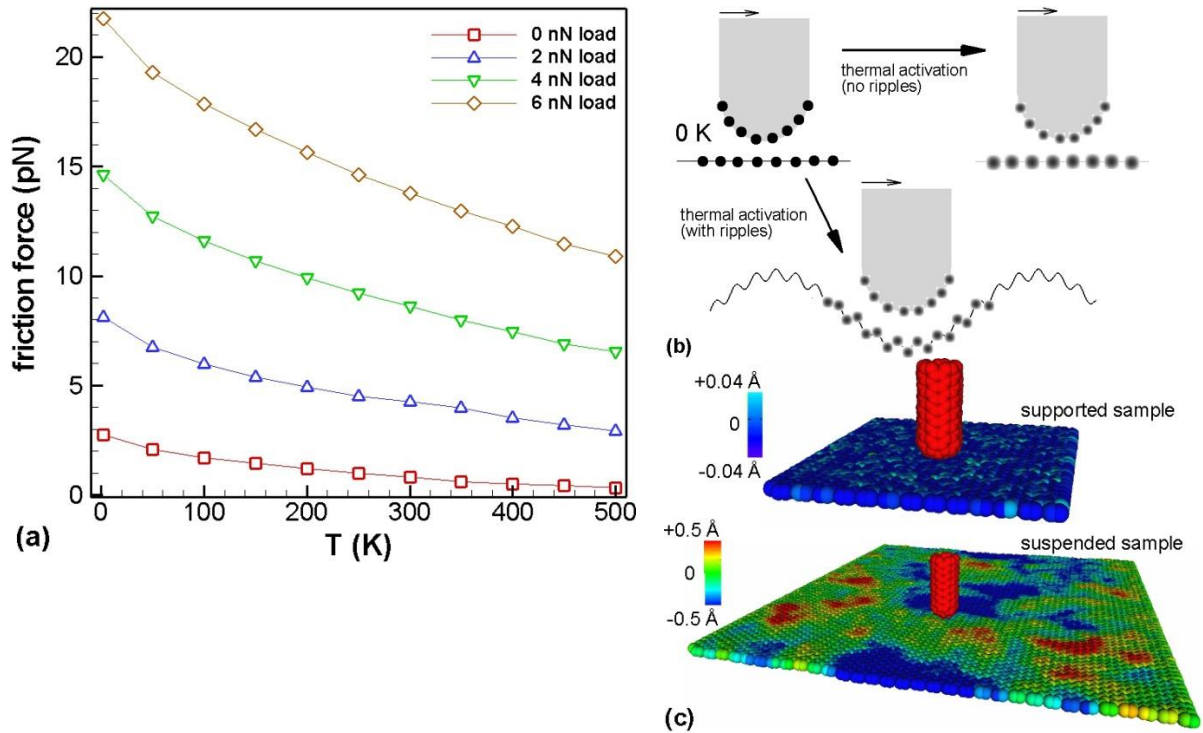

Figure 3. Dependence of friction force on temperature at various normal loads for a supported sample scanned by a $d = 1.2\ nm$ tip (a), schematic representation of the rippling process (b), and typical simulated surface morphology snapshots in supported and suspended samples at 0nN load and T = 300 K (c). The color ranges in (c) correspond to the atomic out-of-plane positions in the corresponding samples.

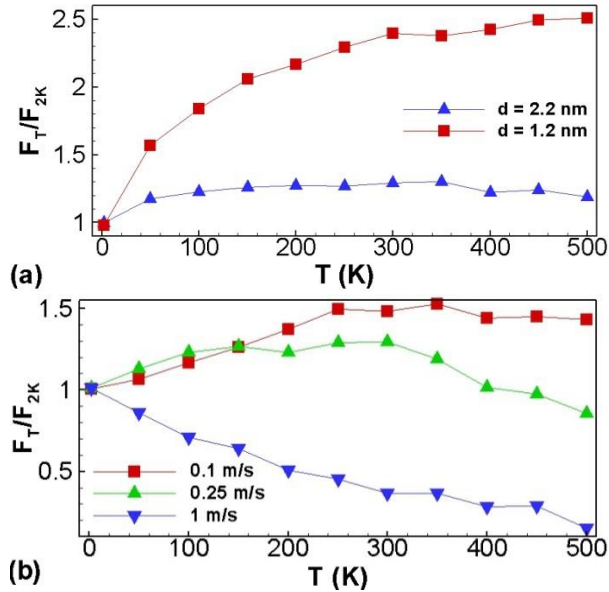

Figure 4. Normalized friction vs. temperature for $d = 2.2\ nm$ and $d = 1.2\ nm$ tips (0 nN load) for the 8192-atom sample (a); Normalized friction as a function of temperature for a 5408-atom sample scanned by a $d = 1.2\ nm$ tip at various scan rates (b).

**Supplemental information for the paper:**
*Effects of thermal rippling on the frictional properties of free-standing graphene*
by A. Smolyanitsky

*S1. Thermally activated friction*

In the theory of thermally activated friction, temperature (and velocity) dependence is expressed via [1]:

$$\frac{1}{\beta k_B T}(F_{max} - F)^{3/2} = \ln\frac{v_0}{v} - \frac{1}{2}\ln\left(1 - \frac{F}{F_{max}}\right), \qquad (S1)$$

where $F_{max}$ is the lateral force amplitude at 0 K, $v_0$ is as defined in the main text with a fitting parameter $f_0 \sim \frac{1}{2\pi}\left(\frac{k}{m_{tip}}\right)^{1/2}$ ($k = \frac{k_0 k_1}{k_0 + k_1}$, where $k_0 = 1.53\ N/m$, and $k_1 = 10\ N/m$ are the first derivative of the lateral force with respect to the sliding distance (Fig. S2) and the virtual cantilever lateral stiffness, respectively) at a given normal load for a tip with an effective mass $m_{tip}$, and $\beta = \frac{3\pi\sqrt{F_{max}}}{2\sqrt{2}a}$ [1]. Note that the average friction force in Eq. S1 is equal to the lateral force amplitude $F$, which assumes overrelaxation of the tip-sample contact. Here (and often experimentally) the actual friction force $F_f$ (lateral force offset, also see section below) is only a fraction of $F$ with $\alpha = F_f/F$ corresponding to the overall dissipative properties of the experimental or simulated system.

For a light CNT tip comprised of approximately 200 carbon atoms, $m_{tip} = 4 \times 10^{-24}\ kg$, which, with $k = 1.5\ N/m$ yields $f_0 = 97.6\ GHz$.

Given the values of $f_0$ and $k$ obtained above, we obtain $v_0 = 2.54\ m/s$. Again, one should note that the value of $k_0$ (and thus $f_0$) in general are not constant, depending on the amplitude of the lateral force as a function of sliding distance (Fig. S2), making these values load-dependent.

We solved Eq. (S1) numerically for $F$ and report here $F_f$, using the lateral force amplitudes and the values of $\alpha$ at 2 K as taken directly from the simulated data; $f_0$ was the only fitting parameter. The results are shown in Fig. S1 for $v = 1\ m/s$. In particular, no increasing trends with respect to temperature are observed in a general sweep of $f_0$ (Fig. S1 (b)) and *no Eq. (S1) solutions were found for $f_0 < 4\ GHz$*.

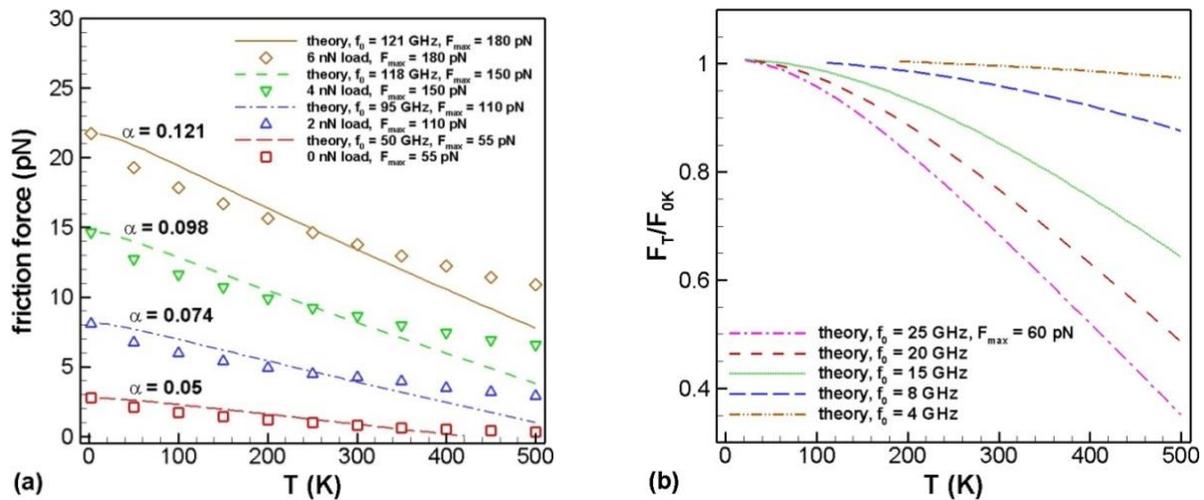

Figure S1. Comparison between simulated data and existing theory for the supported sample at all simulated normal loads (a); a sweep of $f_0$ for a typical $F_{max}$ value obtained from simulations (b).

*S2. Friction force calculation*

The high computational efficiency of the harmonic constraints enabled us to simulate 10 ns of scanning (unless stated otherwise), providing good statistics in terms of the number of lateral stick-slip events experienced by the tip. All force averages were calculated during the last 7.5 ns of each simulation. The mean friction forces presented in the main text were calculated as averages of the lateral force trace data (with a total simulated bandwidth of 250 THz, as dictated by the time-step of 1 fs and the rate of tip-sample force output of every 20 time-steps). A low-pass filter was applied to the raw lateral force data in order to remove added high-frequency thermal noise prior to calculating averages. The effective bandwidth of the filter was 20 GHz, while the characteristic stick-slip frequency corresponded to 4 GHz (at the tip highest sliding velocity of 1 m/s). The grand averages were calculated from the per-bin averages, as described in Fig. S2.

The stick-slip periodicity necessary for combining the data accurately into bins was calculated directly from the Fourier transform peak of the lateral force data, as shown in the inset of Fig. S2. The use of bins was dictated by the fact that the absolute value of the sought average (offset) is about an order of magnitude smaller than the (locally varying) lateral force amplitude (see values of α in Fig. S1 (a)). In addition, the periodicity of the lateral force is distributed over a distance of about 0.15 Å, as shown in the inset of Fig. S2, contributing to the overall variation of local average between the stick-slip events.

Throughout the presented average friction data, the relative standard deviation varied from 1 % at the temperature of 2 K to 13 % at 500 K.

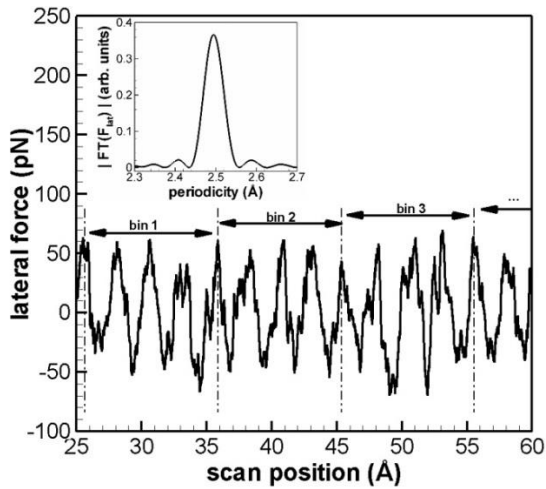

Figure S2. Calculation of average friction from lateral force data (from 0 nN normal load, 300 K of Fig. 2 (a) of main text) with use of data bins. The inset shows the Fourier spectral density of the lateral force data presented; the position of the peak is the effective lattice constant of the scanned sample $\lambda$. The width of each bin is an integer number of $\lambda$. All lateral scans were performed along the *negative* direction of the Y-axis, resulting in a *positive* average.

*S3. Graphene model*

The computationally efficient graphene representation used in this work is an extension of the approach presented [2] and used earlier [3-6], based on a generic model in molecular mechanics [7, 8]. The model was described in detail and dynamically tested in [9] using Nosé-Hoover thermostatics.

For this work, the stiffness constants $k_{bond} = \frac{\partial^2 V_{ij}}{\partial r^2} = 698.13\ N/m$, $k_{ang} = 2\frac{\partial^2 V_{ij}}{\partial \cos\theta^2} = 8.08\ eV$, $k_{dih} = \frac{\partial^2 V_{ij}}{\partial \alpha^2} = 0.360\ eV/rad^2$ [9]. The used model is mathematically guaranteed to yield 0 K structural properties of graphene in agreement with the "parent" bond-order potential for small isotropic strains [2]. In addition to the previously published validation of the approach, we specifically tested our harmonic constraint model against the "parent" optimized second-generation bond-order Brenner potential [10] by directly simulating the thermal rippling process using both methods and calculating the time-averages of the effective ripple height $\langle h^2 \rangle^{1/2}$, thermostatted with use of the Langevin scheme along the sample perimeter, as used in this work. The results of the comparison are shown in Fig. S3, as calculated for a 8192-atom graphene sample with full periodic boundary. The rippling magnitudes (and thus the effects thereof, as reported in the main text) are somewhat *underestimated* by the harmonic constraint model, compared to the optimized Brenner potential, arising primarily from higher-amplitude modes at $q < 0.2$ Å$^{-1}$ ($\lambda > 3.1$ nm) in the latter (see inset). The differences between the two models increase with increasing temperature. Such increasing discrepancy is natural, because agreement is only expected near 0 K, as follows from the constraint energy form used [9]. Nevertheless, the order of magnitude of the simulated rippling heights for both models is qualitatively consistent with the fundamental theory of thermally fluctuating membranes [11], as well as with previously published work [6, 12, 13].

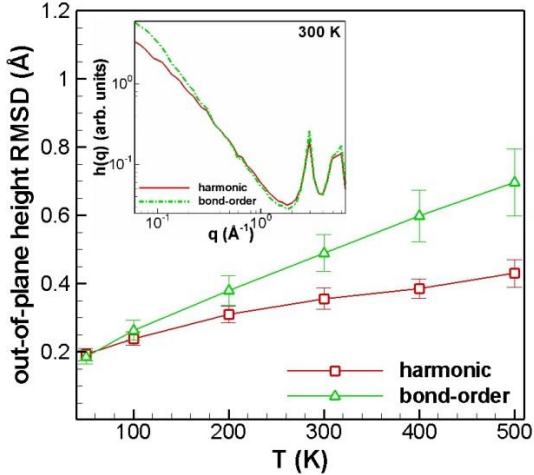

Figure S3. Out-of-plane height RMSD as a function of temperature for the 8192-atom graphene sample. The bars represent the time-variance of the calculated RMSD. The inset shows spatial distributions of the out-of-plane ripples at T = 300 K.

*S4. Scaling and lateral strain*

The effects of scaling are shown in Fig. S4, where in (a) we plot the friction force as a function of temperature for various sample sizes at 0 nN normal load. The differences in $\left(\frac{dF}{dT}\right)$ trends, as well as in the rippling distributions (see inset) are observed, depending on size, demonstrating high sensitivity of $\left(\frac{dF}{dT}\right)$ on the sample size. This indeed includes possible effects of anharmonic coupling, manifested by local decreases in distribution slopes at $q < 0.2$ Å$^{-1}$ ($\lambda > 3.1$ nm), depending on sample size (top left of inset), as mentioned in the main text and consistent with [6]. Additionally, differences in the Bragg peak heights (especially the second peak) are also observed (bottom right of inset), which directly affects the thermal activation mechanism *b(T)*. In Fig. S4 (b), we examine the effect of lateral strain: a $\left(\frac{dF}{dT}\right) > 0$ trend is

S3

observed below 100 K for the strained sample, in contrast with the strain-free case. As shown in the inset, rippling is suppressed overall as a result of strain. Importantly, the local distribution slope is modified significantly throughout $q < 1.0$ Å$^{-1}$ ($\lambda > 0.63$ nm), as expected from pre-stretching a membrane [11]. The first Bragg peak in the inset is also suppressed, again likely contributing to $b(T)$.

*All rippling distributions were averaged over a 1000 sets of Fourier data from atomic position snapshots.*

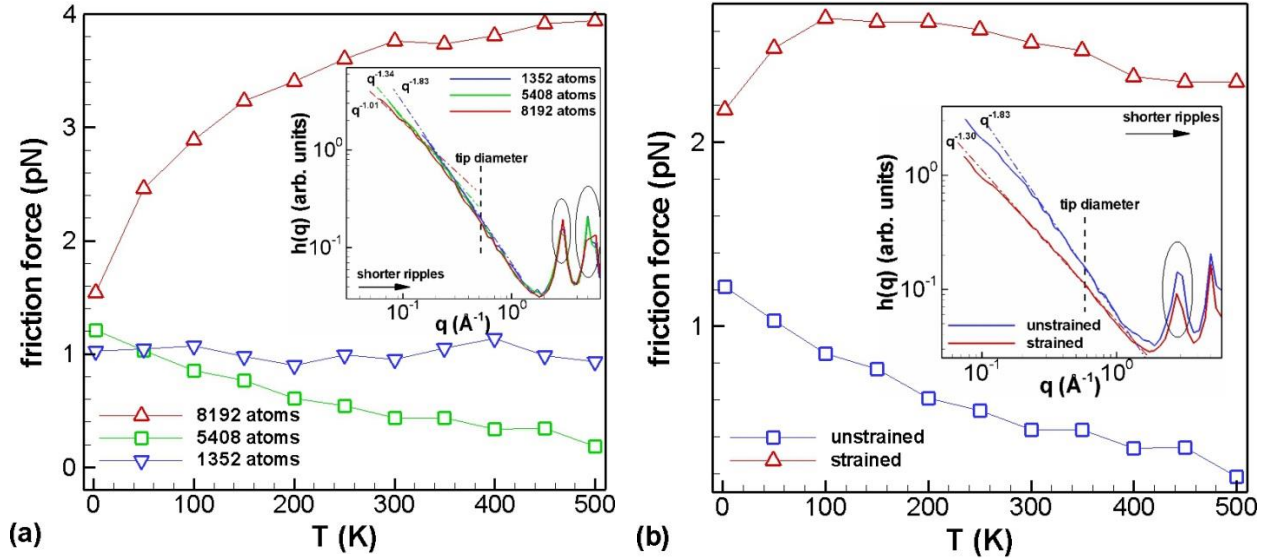

Figure S4. Friction as a function of temperature for different sample sizes at 0 nN normal load; inset shows the $h(q)$ distributions obtained from the snapshots of atomic positions at T = 300 K (a); friction as a function of temperature at 0 nN normal load for a 5408 sample without strain and with 1.3 % lateral strain.

## S5. Simulation and theory comparison

Although the theoretical discussion presented in the main text is qualitative, in Fig. S5 we show the data set from Fig. 4 (a) alongside the corresponding fits of Eq. (5). Here, we assume that the *functional form* of Eq. (S1) is unaffected by the presence of waves in free-standing samples, and thus the (1- $b(T)$) portion of Eq. (5) is solved directly with use of Eq. (S1). The effects of waves on the actual physics (within the assumed Tabor-like model) are then accounted for by the values $f_0$ and ε, used as fitting parameters.

The values of $F_{max}$ for the 1.2 nm and 2.2 nm tip were taken directly from simulation and are equal to 60 pN and 100 pN, respectively. The values of $f_0$ were set to 8 GHz and 20 GHz for the 1.2 nm and 2.2 nm wide tip, respectively. With such values of $f_0$, the $b(T) << 1$ hypothesis for the free-standing samples is indeed supported (see Fig. S1 (b) above). One notes that the increase of $f_0$ for the 2.2 nm wide tip (relative to the 1.2 nm tip) makes qualitative sense, as the lateral stick-slip amplitude increases with the tip diameter. The fitting values of ε are shown in Fig. S5.

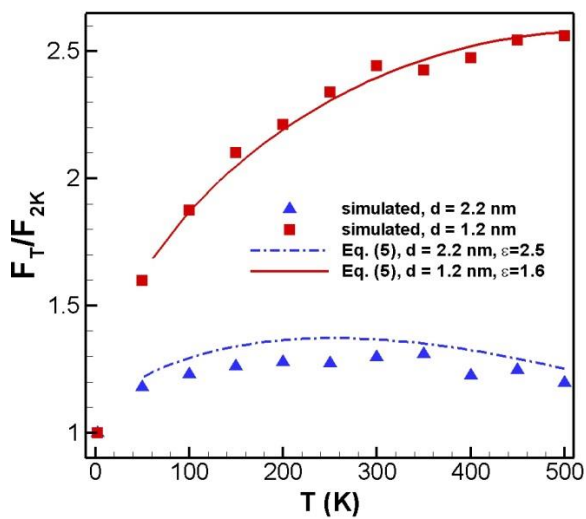

Figure S5. Data from Fig. 4 (a) alongside corresponding Eq. (5) fits. The continuous lines are shown only for temperatures, where the solution of Eq. (S1) was found.